# The Construction of Augmented Designs in Square Arrays

Emlyn WILLIAMS and Hans-Peter PIEPHO

Augmented designs are typically used in early-stage breeding programs to compare single replicates of test entries by combining them with replicated check varieties. One or two dimensional incomplete blocking can be incorporated in the design to accommodate possible site variation. An augmented design in a square array can be derived from a smaller row-column design (the contraction). In a recent paper Bailey and Haines (2025) investigated the link between an augmented design in a square array and its contraction. Here we formally establish this connection by expressing the average efficiency factor of the augmented design in terms of that of its contraction. A consequence of this is that an optimal contraction can be used to construct an optimal augmented design. The table of cyclic contractions presented by Bailey and Haines (2025) is updated in terms of optimality. Specifically, in cases where a cyclic contraction is not optimal, an augmented design with optimal or near-optimal efficiency can be obtained via computer search.

## 1. INTRODUCTION

In plant breeding programs, a large number of entries usually need to be tested in early generations. Typically, the amount of seed available per entry is sufficient for only a single plot. In this situation, estimating an error variance and allowing for local adjustments by blocking can be accommodated by allocating a certain fraction of plots to check varieties that are replicated (Kempton, 1984). One way to generate a design with unreplicated test entries and replicated checks is to employ a blocked design for the check varieties and augment these blocks with test entries. Such designs were first proposed by Federer (1956, 1961) under the label augmented designs. Randomizing test entries and check varieties within each augmented block provides a valid estimate of the error variance and permits adjustments for block effects based on a linear model.

---

E.R Williams, Statistical Support Network, Australian National University, Canberra, Australia (E-mail: emlyn@alphastat.net)

H-P Piepho, Biostatistics Unit, Institute of Crop Science, University of Hohenheim, Stuttgart, Germany

More recently Piepho and Williams (2016) and Vo-Thanh and Piepho (2020) provided methods for the construction of augmented designs in a two-way arrangement of plots and gave references to earlier work. In order to allow for the prospect of both row and column adjustment, it is important that there are adequate numbers of replicated check varieties to provide error degrees of freedom for the comparison of test entries. This can require around 15 – 25% of plots being assigned to check varieties. An alternative is to replicate some of the test entries to generate error degrees of freedom; such designs are called partially replicated (p-rep) designs (Cullis et al., 2006). It is then usual to still include a smaller number of replicated check varieties in the p-rep design; this can have the advantage of allowing a comparison of new material with established check varieties (Williams et al., 2024, section 7.7).

Bailey and Haines (2025) investigated the construction of augmented designs in square arrays from smaller auxiliary designs (contractions). In section 3.2 they derived relationships between the average variance of estimates of the pairwise differences between the check varieties and unreplicated test entries and raised the possibility of a link between average variance quantities for the augmented design and its contraction. As Bailey and Haines (2025) emphasize, this link is important to enable optimality criteria for the contraction to flow through to optimality for the augmented design. In this paper we derive this connection by linking the average efficiency factor for the augmented design to that of its contraction. We do this by identifying a close connection to methodology used by Patterson and Williams (1976) in associating a two-replicate resolvable design to its contraction. Bailey and Haines (2025, Table 4) list a number of cyclic designs which can be used as contractions for the associated augmented designs. Hence, given the algebraic link between the efficiency factors of the contraction and augmented design, it is important to use optimal contractions. Here we tabulate a number of improvements on the cyclic designs listed by Bailey and Haines (2025).

## 2. CONSTRUCTION

We start with a row-column design for $v$ treatments in $v$ columns and $k$ rows and where the rows comprise a complete replicate of the treatments. This design is denoted as the contraction, from which an augmented design will be generated. Federer and Raghavarao (1975) explain how by switching the roles of the $k$ rows and $v$ treatments in the contraction we can generate the positions of $k$ check varieties in the $v \times v$ augmented design. The remaining $v(v-k)$ cells can be filled with unreplicated test entries.

**Example 1.** Suppose for $v = 5$ and $k = 3$, the contraction is given by the row-column design:

| | | Column 1 | 2 | 3 | 4 | 5 |
|---|---|---|---|---|---|---|
| | A | 2 | 4 | 5 | 1 | 3 |
| Row | B | 5 | 3 | 2 | 4 | 1 |
| | C | 1 | 2 | 3 | 5 | 4 |

Switching the roles of the *k* rows and *v* treatments means that the row labels A, B and C in the contraction become the labels for the check varieties in the augmented design, and the treatment labels in the contraction become the row positions of the check varieties. This conversion is applied separately for each column of the contraction, leading to the following location of the check varieties A, B and C in the augmented design, which by construction has the same number of columns as the contraction:

| | | Column 1 | 2 | 3 | 4 | 5 |
|---|---|---|---|---|---|---|
| | 1 | C | | | A | B |
| | 2 | A | C | B | | |
| Row | 3 | | B | C | | A |
| | 4 | | A | | B | C |
| | 5 | B | | A | C | |

The blank spaces are then randomly filled with the single replicates of 10 test entries to complete the augmented design.

Patterson and Williams (1976) presented methodology to demonstrate how a $k \times v$ contraction can be used to construct a two-replicate resolvable incomplete block design for *vk* treatments with *v* blocks of size *k* in each replicate. They showed that there is a link between the average efficiency factor of the contraction ($E_{con}$) and that for the resolvable design, where the average efficiency factor of an incomplete block design is defined as the harmonic mean of the canonical efficiency factors (John and Williams, 1995, section 2.3). In particular, Patterson and Williams (1976) derived the relationship

$$E_{res} = \frac{(vk - 1)}{(vk - 2v + 1) + 4(v - 1)/E_{con}} \qquad (1)$$

where $E_{res}$ is the average efficiency factor of the resolvable design. John and Williams (1995, section 4.7) provided more detail on the construction process. Note that although the contraction for both the resolvable and augmented designs has been specified as a row-column design, because the rows comprise a complete replicate of the treatments, the average efficiency factor of the row-column contraction is the same as that just considering the columns as incomplete blocks.

In a similar vein we can use Theorem 7 of Patterson and Williams (1976) to develop the following link between the average efficiency factor for the augmented design ($E_{aug}$) and that for the contraction:

$$E_{aug} = \frac{(v^* - 1)}{(v^* - 2v + 1) + 2v(v-1)/(kE_{con})} \qquad (2)$$

where $v^* = v \times v - k \times (v - 1)$ is the total number of treatments in the augmented design. A derivation of (2) is provided in the Appendix.

Expression (2) defines the average efficiency factor for the whole design, including the check varieties. But because these are orthogonal to both rows and columns in the augmented design with average efficiency factor of one, we can simply subtract the quantity $k$ from numerator and denominator to obtain the average efficiency factor ($E_{test}$) of the test entries, i.e.

$$E_{test} = \frac{(v^* - 1) - k}{(v^* - 2v + 1) - k + 2v(v-1)/(kE_{con})} \qquad (3)$$

Furthermore the average pairwise variance for the test entries ($A_{test}$) is given by $A_{test} = {}^{2}/_{E_{test}}$, e.g. see Williams and Piepho (2015), equation (3).

**Example 2**. Bailey and Haines (2025, Example 5.1) discuss a contraction for 12 treatments in 12 incomplete blocks of size 3. This is an optimal incomplete block design with $E_{con} = 0.68006$ thus giving $A_{test} = 4.0075$, which agrees with the result in their Table 6.

## 3. OPTIMAL CONTRACTIONS

Bailey and Haines (2025) have used tables of cyclic designs from Lamacraft and Hall (1982) to provide a list of available contractions within useful parameter ranges (see their Table 4). But for many combinations of *v* and *k*, cyclic designs do not provide optimal contractions and hence from (2) and (3), will not result in optimal augmented designs in square arrays. To illustrate this we have taken the $k = 3$ and 4 examples from Table 4 of Bailey and Haines (2025) and obtained the corresponding initial cyclic design $E_{con}$ values from Lamacraft and Hall (1982). We have then generated optimal or near-optimal contractions for the same parameters using the design generation package CycDesigN (VSNi, 2025) and recorded their $E_{con}$ values; results are presented in Table 1. For 20 of the 24 designs in this table, the computer-generated designs have higher $E_{con}$ values when compared with the cyclic contractions and thus would be preferred; for the other four designs, the cyclic contractions (in bold) have the same $E_{con}$. Also included in Table 1 is the percentage of a theoretical upper bound for $E_{con}$ achieved by the computer-generated designs. Details on the derivation of upper bounds are given by John and Williams (1995, section 2.8).

Bailey and Haines (2025) also present a small table of non-cyclic contractions (their Table 6). In Table 2 we convert their $A_{abd}$ values to the corresponding $E_{con}$ using the relationship $E_{con} = {}^{2}/_{(kA_{abd})}$ (Williams and Piepho, 2015). Here four of the five examples concur with those generated by CycDesigN.

Table 1. Comparison of $E_{con}$ for cyclic and computer-generated contractions and percentage of a theoretical upper bound for the latter.

| | | Cyclic | CycDesigN | |
|---|---|---|---|---|
| *v* | *k* | *Econ* | *Econ* | % UB |
| 10 | 3 | 0.6998 | 0.705895 | 99.41 |
| 11 | 3 | 0.6762 | 0.690163 | 99.13 |
| 12 | 3 | 0.6726 | 0.680062 | 99.30 |
| 13 | 3 | 0.6667 | 0.669481 | 99.14 |
| 14 | 3 | 0.6527 | 0.663024 | 99.36 |
| 15 | 3 | 0.6409 | 0.660377 | 100.00 |
| 16 | 3 | 0.6321 | 0.647969 | 99.32 |
| 17 | 3 | 0.6222 | 0.643898 | 99.78 |
| 18 | 3 | 0.6116 | 0.637262 | 99.73 |
| 19 | 3 | 0.6111 | 0.631561 | 99.72 |
| 20 | 3 | 0.6011 | 0.627431 | 99.87 |
| 14 | 4 | **0.8029** | 0.802941 | 100.00 |
| 15 | 4 | **0.7955** | 0.795455 | 100.00 |
| 16 | 4 | 0.7872 | 0.789474 | 100.00 |
| 17 | 4 | 0.7803 | 0.782335 | 99.84 |
| 18 | 4 | 0.7762 | 0.777101 | 99.83 |
| 19 | 4 | **0.7725** | 0.7725 | 99.84 |
| 20 | 4 | **0.7686** | 0.768571 | 99.87 |
| 21 | 4 | 0.7624 | 0.765069 | 99.90 |
| 22 | 4 | 0.7599 | 0.761053 | 99.83 |
| 23 | 4 | 0.7552 | 0.758038 | 99.84 |
| 24 | 4 | 0.7533 | 0.754688 | 99.77 |
| 25 | 4 | 0.749 | 0.751914 | 99.75 |
| 26 | 4 | 0.7454 | 0.749165 | 99.71 |

Table 2. Comparison of $E_{con}$ for non-cyclic and computer-generated contractions and percentage of a theoretical upper bound for the latter.

| | | Non-Cyclic | | CycDesigN | |
|---|---|---|---|---|---|
| $v$ | $k$ | $A_{abd}$ | $E_{con}$ | $E_{con}$ | % UB |
| 9 | 3 | 0.9167 | 0.7273 | 0.727273 | 100.00 |
| 10 | 3 | 0.95 | 0.7018 | 0.705895 | 99.41 |
| 12 | 3 | 0.9803 | 0.6801 | 0.680062 | 99.30 |
| 16 | 4 | 0.6333 | 0.7895 | 0.789474 | 100.00 |
| 25 | 5 | 0.4833 | 0.8276 | 0.827586 | 100.00 |

## 4. DISCUSSION

An important consequence of (2) is that the best augmented design in a square array will be obtained from the best available contraction. For some parameter combinations, optimal contractions can be obtained from existing construction methods. In addition, the use computer design generation packages such as CycDesigN (VSNi, 2025) can be very effective, especially as the contractions of practical interest for the generation of such augmented designs are likely to be quite small (e.g. $k \leq 4$). In such cases, the powerful computers and algorithms of today can often produce optimal designs very quickly, as can be indicated by tight upper bounds on the average efficiency factor (John and Williams, 1995, section 2.8). These upper bounds are theoretically derived and can either correspond to the optimal design achievable or are as tight as the current development can provide. In the latter case it is hence likely that some of the designs that do not achieve 100% of the upper bound, are in fact optimal. In any case the upper bounds provide a valuable ceiling for determining a stopping point for the iterative improvement of the average efficiency factor via computer search. CycDesigN displays these upper bounds and for practical purposes, we usually suggest that if, after the iterative improvement slows down, a generated design is within 1-2% of the upper bound, then the user can terminate the search process; this especially the case with larger experiments such as breeding trials.

Finally, a topic of further research could be an investigation of whether the idea of contractions can also be used for augmented designs in rectangular arrays of plots; and even whether the construction of efficient p-rep designs including check varieties can also be produced.

## APPENDIX

### Derivation of equation (2)

For a $v \times v$ array with $k$ check varieties per row and column, we assume the model

$$y = X\tau + Z_R\rho + Z_C\gamma + \varepsilon$$

where $X$ is $n \times v^*$, $Z_R$ and $Z_C$ are $n \times v$ with $n = v \times v$.

The information matrix is given by

$$A = r^\delta - \tfrac{1}{v}N_R N_R' - \tfrac{1}{v}N_C N_C' + \tfrac{1}{v^2}rr' \text{ (John and Williams, 1995, eq 5.3)}$$

where $r$ is a vector of treatment replications, in order with check varieties at the end, i.e.

$r' = (1 \dots\dots 1, v \dots v)$, $N_R = X'Z_R$, $N_C = X'Z_C$ and $r^\delta = diag(r)$.

We are interested in the canonical efficiency factors (*cefs*) of $A^* = r^{-\delta/2} A r^{-\delta/2}$ (John and Williams, 1995, eq 5.5) which we can write as

$$A^* = I_{v^*} - r^{-\delta/2}\tfrac{1}{v}X'[Z_R Z_C]\begin{bmatrix} Z_R' \\ Z_C' \end{bmatrix} X r^{-\delta/2} + \tfrac{1}{v}J_{v^*}$$

where $I_{v^*}$ is the identity matrix of size $v^*$and $J_{v^*}$ is the $v^* \times v^*$ matrix of ones. The $v^* - 1$ non-trivial eigen values of the central term will the same as those of

$$\tfrac{1}{v}\begin{bmatrix} Z_R' \\ Z_C' \end{bmatrix} X r^{-\delta} X'[Z_R Z_C] \qquad \text{(A1)}$$

along with $v^* - 2v - 1$ zero eigen values.

The structure of the $v \times v$ array allows us to calculate the components of (A1). For example the two diagonal $v \times v$ blocks both have the form

$$\tfrac{1}{v}\left\{(v-k)I_v + \tfrac{k}{v}J_v\right\}$$

The off-diagonal blocks are

$$\tfrac{1}{v}\left\{\tfrac{k}{v}J_v + Q\right\} \text{ and } \tfrac{1}{v}\left\{\tfrac{k}{v}J_v + Q'\right\}$$

where $Q$ is a $v \times v$ matrix with $(i,j)th$ element equal to 1 if the augmented design contains a test entry in the $(i,j)th$ position, $i,j = 1 \dots v$, hence

$$Q1_v = (v-k)1_v.$$

We note that an eigen vector of (A1) is also an eigen vector of

$$B = \tfrac{1}{v}\begin{bmatrix} kI_v & -N \\ -N' & kI_v \end{bmatrix} \qquad \text{(A2)}$$

where $N = Q - J_v$. Hence the $2(v-1)$ non-trivial eigen values of (A2) are eigen values of $A^*$.

We can now recognize $N$ as the incidence matrix of the contraction and hence, using the results of Theorem 7 of Patterson and Williams (1976), we have that if $\theta$ is a *cef* of the contraction, then $\left(1 \mp \sqrt{1-\theta}\right)/2$ are eigen values of $vB/(2k)$.

The average efficiency factor ($E_{aug}$) of the augmented design is the harmonic mean of its *cefs*, hence looking at the sum

$$\frac{2}{\left(1+\sqrt{1-\theta}\right)} + \frac{2}{\left(1-\sqrt{1-\theta}\right)} = \frac{4}{\theta}$$

we have that the sum of the reciprocals of the $2(v-1)$ non-trivial eigen values of $B$ add to

$$\frac{v}{2k}\frac{4(v-1)}{E_{con}}.$$

Combining this sum with $v^{*} - 2v + 1$ unit *cefs* completes the derivation.